# Electrostatic Tactile Display without Insulating Layer

Hiroyuki Kajimoto[1]

[1] Department of Informatics, The University of Electro-Communications, Tokyo, JAPAN

(Email: kajimoto@kaji-lab.jp)

**Abstract ---** This paper explores an approach to eliminating the surface insulating layer in electrostatic (electroadhesion) tactile displays. Electrostatic tactile displays modulate the surface friction by an electrical charge between the skin and the display. Traditionally, the non-conductive dielectric layer has been considered crucial for charge accumulation, as well as for safety to prevent DC current stimulation. However, by utilizing a current control technology for electrotactile displays, we can achieve electrostatic tactile display without the insulating layer. The electrical charge is possibly accumulated in the skin itself or in the air gap between the skin and the electrodes. Safety is maintained by balancing positive and negative current pulses. Furthermore, this system is compatible with existing electrotactile displays. This paper details the system configuration, presentation algorithm, and experimental results. The preliminary trial revealed that five out of eight participants could clearly feel the vibration, confirmed by acceleration recording, while the remaining participants could not experience the sensation.

## 1 Introduction

When operating a smartphone or tablet with a finger, it is important to provide tactile feedback to that finger. One widely studied method for this purpose is the electrostatic (electroadhesion) tactile display [1][2][3][4][5][6][7], which consists of a transparent electrode on the screen and a transparent insulating (non-conductive) layer on it. The electrode is typically participanted to a frequency-modulated high voltage, which, when touched by the user through the insulating layer, produces an electrical adhesive force. The force can generate a frictional sensation when the user slides his/her finger on the surface.

The insulating layer described above is considered essential for electrostatic tactile displays, with two main roles. First, it has the safety role of preventing direct current from flowing into the skin. Second, it serves as a place for charge storage. Because of the second role, thin materials with high dielectric constant have often been used.

The argument of this study is that this insulating layer is not always necessary, and bare electrodes can serve as an electrostatic tactile display (Fig.1 ). Since the skin has a stratum corneum that is expressed as a pair of capacitance and resistance, and a layer of air gap exists when the finger is traced, it is considered possible to store the electric charge on these layers. To address the safety concern of direct current flow, we can provide positive and negative current pulses and balance them to prevent the continuous accumulation of electric charge. This means that the electrotactile display technology [8][9], another common method of providing tactile sensations by direct nerve stimulation, can be utilized.

The same device can also work as an electrotactile display, making it possible to use "electrotactile display when the finger is not moved" and "electrostatic tactile display when the finger is moved". There have been attempts to combine electrotactile and electrostatic displays [10], but they needed to prepare two types of electrodes, with and without the insulating layer.

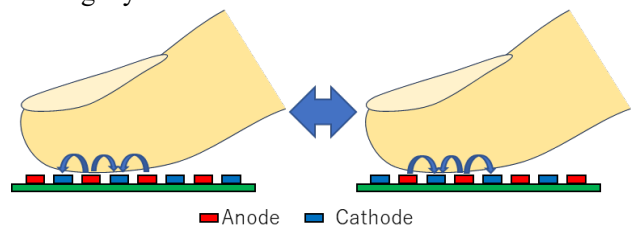

Fig.1 Electrostatic (electroadhesion) tactile display without insulating layer is achieved by continuous biphasic current stimulation.

In our previous report [11], we have shown that both electrostatic and electrotactile tactile sensations can be realized with a relatively low-voltage electrical stimulator and bare electrodes, inspired by previously proposed electrotactile algorithm [12]. However, due to the low voltage, the electrostatic stimulation was too weak to be felt by some participants.

This paper describes the hardware and driving algorithm to build an electrostatic tactile display, which is based on the conventional electrotactile display system. We also show the measurement results of physical vibrations generated by tracing a finger.

## 2 SYSTEM AND ALGORITHM

### 2.1 Hardware

Fig.2 describes the system configuration. The circuit is basically the same as that of the electrical stimulator already reported [13], which consists of a microcontroller, a high-voltage generator circuit (up to 300 V), a DA converter, an AD converter, a voltage/current converter circuit, and a group of switch-pairs. The switch-pair consists of one upper-side switch and one lower-side switch per electrode. This means that when the upper switch is turned on, the electrode becomes a current source, and when the lower switch is turned on, the electrode becomes ground.

When we use this device as an *electrotactile* display, only one stimulating electrode is connected to the current source and all the other electrodes are connected to ground. In this case, the stimulating electrode becomes the anode and all surrounding electrodes become cathodes, resulting in anodic stimulation. Conversely, if only the stimulating electrode is connected to ground and all other electrodes are connected to a current source, cathodic stimulation occurs.

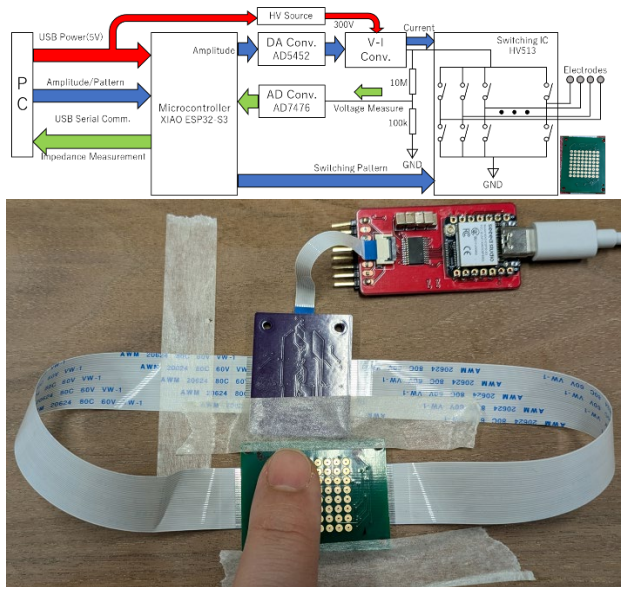

Fig.2 Configuration of the system.

The electrodes used in this study are gold-coated, 2 mm in diameter, 3 mm in distance between electrode centers, and arranged in an 8 × 8 configuration. The entire electrode area is 23 mm × 23 mm, and a finger tracing motion can be performed on the electrodes to generate vibrations caused by electrostatic force.

### 2.2 Algorithm

To generate electrostatic force, the electrodes are driven in a different way than for electrotactile display (Fig.3 ). First, half of the electrodes are connected to the current source, and the rest are connected to the ground (in the figure, electrode rows alternately connected to current source and ground, but for example, they can be a checker board pattern where the black parts are the current source and the white parts are the ground). Then, 50 us square current pulse is applied from the current source.

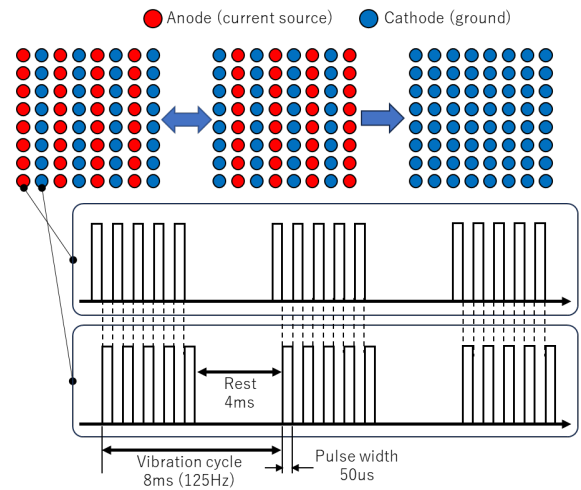

Fig.3 Rendering method for 125 Hz vibration.

Next, the roles of the electrodes are swapped. The group of electrodes that was connected to the current source is connected to ground, and the group of electrodes that was connected to ground is connected to the current source, and 50us square current pulse is applied. This balances the electric charges (the amount of charge flowing out and in is equal, and no charge continues to accumulate in the skin).

This switching is repeated many times. In the example shown in the figure, it is repeated for 40 times during 4 ms period (4 ms / (50us * 2) = 40). During this 4 ms period, electric charge continues to flow in and out of the skin, creating the electrostatic force necessary for tactile sensation. Then, a pause period of 4 ms is set. This 8-ms repetition gives vibrations of 125 Hz.

Since a high stimulation current causes tactile sensation due to electrical nerve stimulation, the stimulation current must be kept at a level that does not cause such sensation.

## 3 EXPERIMENT

### 3.1 Setup

The purpose of the experiment is to confirm that the proposed method actually produces mechanical vibrations caused by electrostatic force. An accelerometer (BMX055, BOSCH) was mounted on the participants' fingernail, and acceleration during finger tracing movement was recorded.

The participants were first instructed moving their dominant hands' index fingers on the electrode. The current value was set to 10mA (the voltage is automatically regulated but typically around 100 to 200 V). They were instructed to make a circular motion on the electrode at a speed of about two circles per second. Acceleration was recorded for 1 second for each case with and without stimulation.

We used only z axis acceleration (normal to the skin surface). The recorded data were subtracted by moving average of 0.2-second to remove effects by finger orientation.

We recruited eight participants from the laboratory including the author (all males, ages 22-48).

### 3.2 Result

Among the eight participants, three participants could not feel any vibration from the surface. The acceleration measurement also did not give any difference. The remaining five participants felt a stable vibratory sensation.

Fig.4 shows the acceleration data for those five participants. The left column shows when the stimulation was conducted, and the right column shows when there was no stimulation. The average RMS (root mean square) for the stimulation condition was 0.241 G, and that for the no stimulation condition was 0.074 G. Therefore, although the electrostatic tactile display did not work for some participants, it surely gave a strong mechanical vibration to the other participants.

While the cause of large difference among participants was not fully explored, it is highly plausible that the skin moisture played an important role. Participants with soft and moist skin tended not to perceive the electrostatic vibration, although they wiped their fingers thoroughly. The default friction of the electrode substrate might also play an important role, since participants could not move their fingers on the plate smoothly.

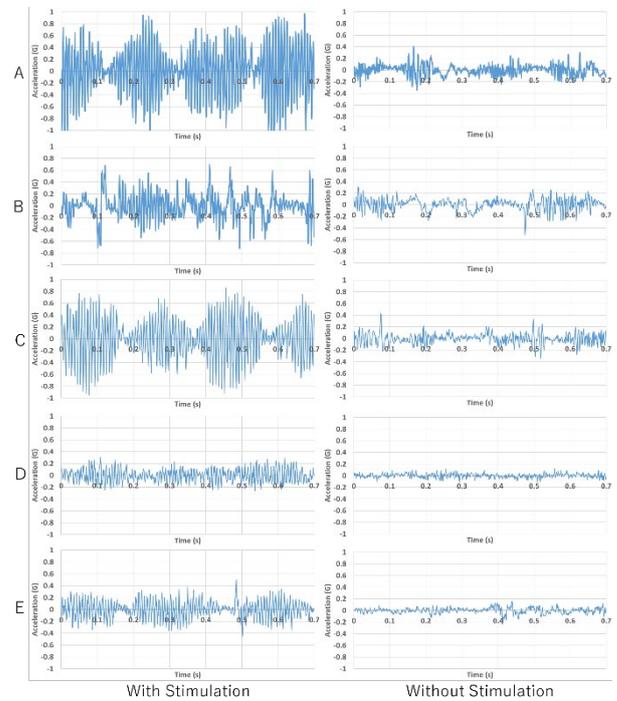

Fig.4 Acceleration for the five participants who felt the electrostatic vibration, measured on the nail during finger tracing motion.

## 4 CONCLUSION

The purpose of this study was to show that electrostatic tactile display can be realized with bare electrodes that are not covered with an insulating layer. Using a conventional electrotactile display, we developed an algorithm for stimulating 8 × 8 electrodes with a waveform that rapidly switches between current source and ground to balance the electric charge. Experimental results showed that for five out of eight participants, the electrostatic stimulation produced mechanical vibration, but we also found that some participants could not feel the vibration.

Future prospects include establishing a more robust stimulation, considering the impedance of individual skin. Furthermore, combination of electrostatic and electrical stimulation needs to be explored.

ACKNOWLEDGEMENT

This work is supported in part by JSPS KAKENHI

Grant NumberJP24K21321.

REFERENCES

[1] Basdogan, C., Giraud, F., Levesque V., Choi,S.: A Review of Surface Haptics: Enabling Tactile Effects on Touch Surfaces, *IEEE Trans. Haptics*, 13(3) 450-470 (2020)

[2] Bau, O., Poupyrev, I., Israr, A., Harrison, C.: TeslaTouch: Electrovibration for touch surfaces, *ACM symposium on User interface software and technology*, pp.283–292 (2010)

[3] Vardar, Y., Guclu, B., Basdogan, C.: Effect of waveform on tactile perception by electrovibration displayed on touch screens, *IEEE Trans. Haptics*, 10(4) 488–499 (2017)

[4] AliAbbasi, E., Sormoli, M.A., Basdogan, C.: Frequency-dependent behavior of electrostatic forces between human finger and touch screen under electroadhesion, *IEEE Trans. Haptics*, 15(2) 416-428 (2022)

[5] Ilkhani, G., Samur, E.: Creating multi-touch haptic feedback on an electrostatic tactile display, *Proc. IEEE Haptics Symposium*: 163-168 (2018)

[6] Osgouei, B.H.: Electrostatic friction displays to enhance touchscreen experience, Modern Applications of Electrostatics and Dielectrics. *IntechOpen* (2020)

[7] Bach-y-Rita, P. B., Kaczmarek, K. A., Michell, T. E., Garcia, A. J.: Form perception with a 49-point electrotactile stimulus array on the tongue, J. *Rehabilitation Research Development*, 35:427–430 (1998)

[8] Kajimoto, H.: Electro-tactile display: principle and hardware. In *Pervasive Haptics, Science, Design, and Application*, Springer Japan, pp.79-96 (2016)

[9] Yem, V., Kajimoto, H.: Comparative evaluation of tactile sensation by electrical and mechanical stimulation, *IEEE Trans. Haptics*, 10(1): 130-134 (2017)

[10] Komurasaki, S., Kajimoto, H., Shimokawa, F., Ishizuka, H.: Characterization of an electrode-type tactile display using electrical and electrostatic friction stimuli, *Micromachines*, 12(3):313 (2021)

[11] Kajimoto, H: Electrostatic and Electrotactile Presentation Using Low Voltage Electrical Stimulation, *IEEE Haptics Symposium*, Work-in-Progress (2024)

[12] Lin, W. et al.: Super-resolution wearable electrotactile rendering system, Sci. Adv.8,eabp8738 (2022)

[13] Kajimoto, H.: Electro-tactile display kit for fingertip, *IEEE World Haptics Conference*, Work-in-Progress, p.587 (2021)